# AI-Driven Design of poly(ethylene terephthalate)-replacement copolymers

Chiho Kim[1,2*], Wei Xiong[1], Akhlak Mahmood[2], Rampi Ramprasad[1,2], Huan Tran[1,2]

1. *School of Materials Science and Engineering, Georgia Institute of Technology, Atlanta, Georgia 30332, USA*

2. *Matmerize, Inc., Atlanta, Georgia 30308, USA*

*E-mail: chiho.kim@gatech.edu

Poly(ethylene terephthalate) (PET), a widely used thermoplastic in packaging, textiles, and engineering applications, is valued for its strength, clarity, and chemical resistance. Increasing environmental impact concerns and regulatory pressures drive the search for alternatives with comparable or superior performance. We present an AI-driven polymer design pipeline employing virtual forward synthesis (VFS) to generate PET-replacement copolymers. Inspired by the esterification route of PET synthesis, we systematically combined a down-selected set of Toxic Substances Control Act (TSCA)-listed monomers to create 12,100 PET-like polymers. Machine learning models predicted glass transition temperature (Tg), bandgap, and tendency to crystallize, for all designs. Multi-objective screening identified 1,108 candidates predicted to match or exceed PET in Tg and bandgap, including the "rediscovery" of other known commercial PET-alternate polymers (e.g., PETG, Tritan, Ecozen) that provide retrospective validation of our design pipeline, demonstrating a capability to rapidly design experimentally feasible polymers at a scale. Furthermore, selected, entirely new (previously unknown) candidates designed here have been synthesized and characterized, providing a definitive validation of the design framework.

## I. Introduction

Designing synthetic polymers has been a central focus of polymer and plastics R&D in both industry and academia for over a century, driving the development of materials with targeted properties for diverse applications.[1] Landmark examples such as Kevlar, Teflon, and Nylon, having originated from laboratory innovations reached commercialization through extensive optimization, validation, and feasibility testing.[2-4] Recent global regulatory efforts including the push for PFAS-free materials,[5] the U.S. Toxic Substances Control Act (TSCA),[6] and the European REACH regulation, have made the development of new polymers increasingly complex.[7] These initiatives demand materials that are non-toxic, sustainable, and environmentally friendly, further constraining the allowable chemical space.

Traditionally, polymer discovery has relied on trial-and-error experiments, which are time consuming and expensive. The vastness of the chemical space, including combinations of homopolymers and copolymers, neat resins and composites, as well as formulation and processing variables, makes efficient exploration and optimization challenging. To address these limitations, artificial intelligence and machine learning (AI/ML) have been actively adopted.[8-15] These approaches connect data collection and curation, numerical representation of polymer structures (fingerprinting),[14,15] model training,[11] and property prediction to guide experimental pathway. While early efforts focused on building predictive models and robust fingerprinting schemes, current trends are shifting toward AI-driven generation and design of new candidate materials.[16-20]

This work presents a polymer design pipeline that employs virtual forward synthesis (VFS)[19] within an AI-assisted design pipeline to craft alternative copolymers for poly(ethylene terephthalate) (PET), one of the widely used thermoplastics in the polyester family. The PET is commonly found in transparent and thermally resistant plastics used for bottles and packaging. In this use case, the design objective is to identify candidate copolymers that exceed PET's performance in key properties, specifically, the target criteria include the glass transition temperature (Tg) higher than PET's Tg range of 65°C to 81°C,[21] and a bandgap as a proxy for optical transparency greater than PET's measured bandgap 3.9 eV.[22] The transparency is also associated with amorphous phase in PET as the crystallinity tends to increase light scattering.[23] Thus, the tendency to crystallization is used as a third design objective. The goal is to identify materials with lower tendency to crystallize than PET, which typically exhibits a crystallinity of less than 40%. (Table I) Two selected candidate copolymers were successfully synthesized and experimentally tested, confirming their predicted properties and demonstrating the practical feasibility of the proposed design approach.

**TBALE I.** Target criteria for design PET-replacement polymers

| Target property | PET, experimental (predicted) | Design goal, predicted |
| --- | --- | --- |
| Glass transition temperature | 65°C - 81°C (72°C) | Greater than 72°C |
| Bandgap | 3.9 eV (3.7 eV) | Greater than 3.7 eV |
| Tendency to crystallize | Less than 40% (40%) | Less than 40 % |

**II. Methods**

All procedures described in this study, including 1) training predictive models for Tg, bandgap, and tendency to crystallize, 2) generation of PET-co-(PET-like) copolymer designs through VFS, 3) multi-objective screening based on target criteria, 4) experimental validation of polymers selected based on synthetic accessibility (SA) score,[24] and 5) feature importance analysis and correlation analysis using SHapley Additive exPlanations (SHAP)[25] analysis and principal component analysis (PCA),[26] were carried out using PolymRize™ platform,[27] a standardized polymer informatics software.

Three predictive ML models were developed to predict Tg, bandgap, and tendency to crystallize for new polymer designs. The Tg model was trained on 8,962 experimentally measured values from a diverse set of homo and copolymers. The bandgap model was trained on 562 density functional theory (DFT)-calculated values using atomistic homopolymer models packed in 3D supercells.[28] The tendency to crystallize model was developed using 111 experimental data points, supported by 432 computationally derived values (based on heat of formation) within a multi-target co-learning framework. All models employed Gaussian process regression (GPR)[29] with radial basis function[30] and white noise kernels, using Polymer Genome fingerprinting scheme, implemented in PolymRize™, to numerically represent the generated copolymers. Model performance on test sets averaged over 5-fold cross-validation (CV) showed root mean square error (RMSE) of 31°C for Tg, 0.6 eV for bandgap, and 12% for tendency to crystallize. While the crystallization model is less accurate than the others, it provides useful guidance for screening.

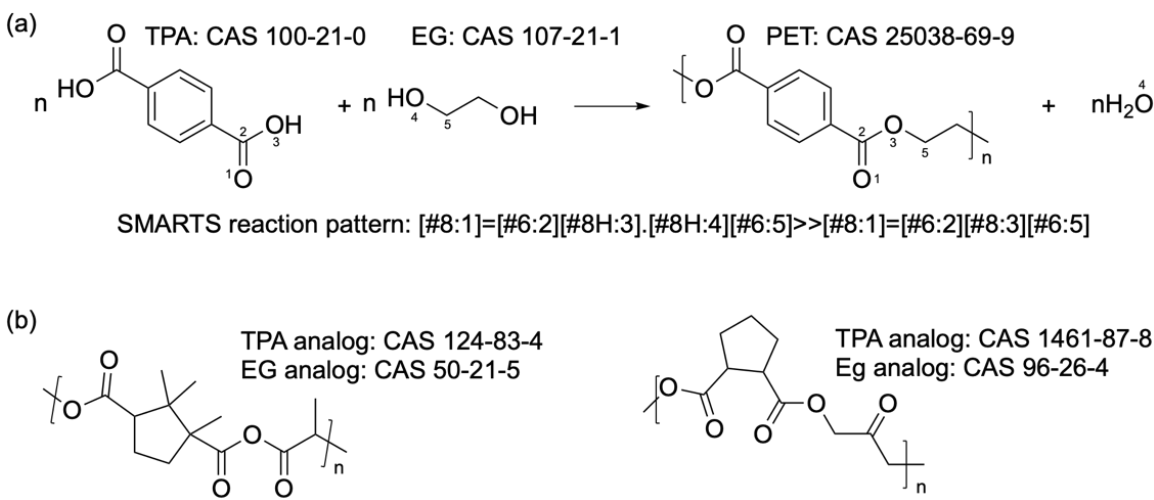

**FIG. 1.** (a) Reaction of TPA and EG to produce PET, and corresponding SMARTS representing condensation of groups - carboxyl from TPA and hydroxyl from EG. (b) Example PET-like polymers generated using the same reaction mechanism but different monomer reactants.

To apply VFS for generating synthetically feasible polymers structurally similar to PET, it is critical to understand the underlying reaction chemistry. Two well-established reaction mechanisms are 1) esterification between terephthalic acid (TPA) and ethylene glycol (EG), and 2) transesterification between dimethyl terephthalate (DMT) and EG. This study focuses on the esterification route, which is a straightforward and classical polycondensation reaction, as illustrated in Fig. 1(a). Candidate monomers for TPA and EG analogs were down-selected from the TSCA inventory of 28,289 registered molecules based on the structural criteria essential for esterification. For TPA replacements, 72 monomers were selected based on the presence of two terminal -COOH groups and five- or six-membered carbon rings, either aromatic or aliphatic. For EG replacements, 168 monomers were chosen using the following criteria: terminal -OH groups, composed solely of carbon and hydrogen functional groups, and a maximum of eight carbon atoms to limit the size and feasibility of the generated products. Through combinatorial pairing of these 72 TPA-like and 168 EG-like monomers, 12,100 candidate PET-like polymers were generated via esterification. Two example structures, along with the CAS numbers of their monomer components, are shown in Fig. 1(b). By combining the original PET repeat unit with each PET-like polymer in a 1:1 molar ratio, a set of 12,100 PET-co-(PET-like) copolymers was constructed using the RxnChainer feature of PolymRize,[31] and prepared for property prediction.

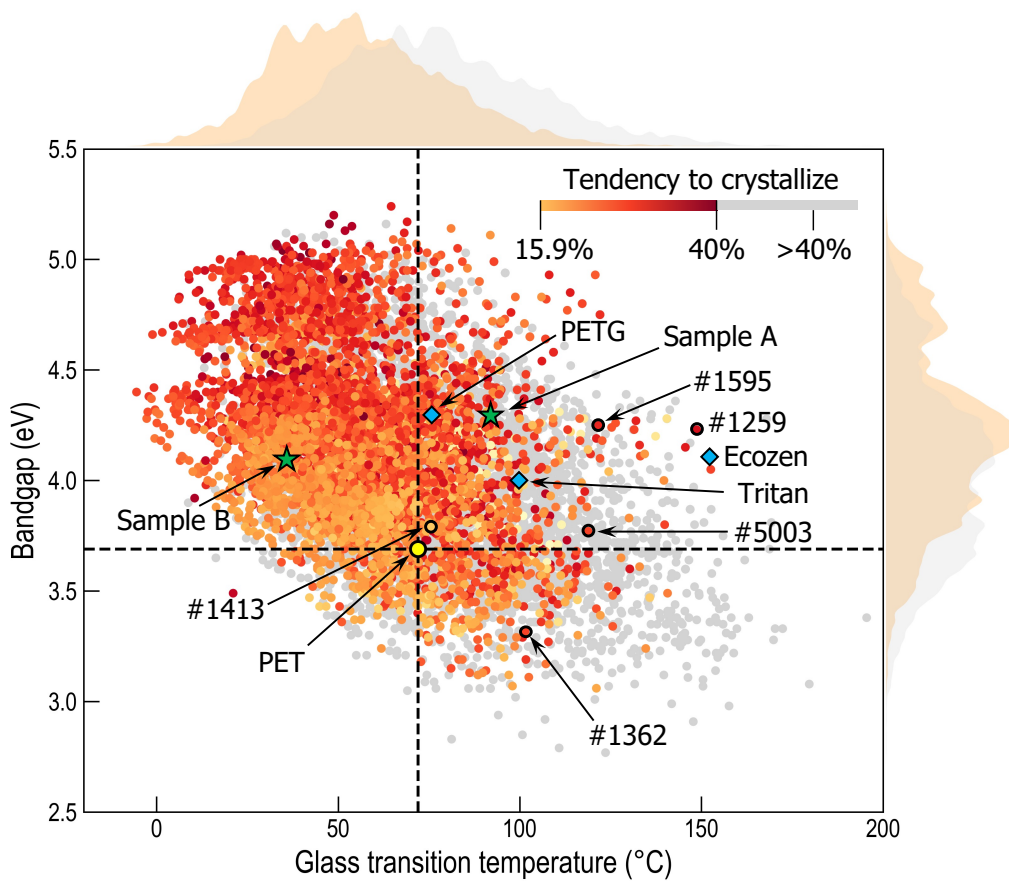

**FIG. 2.** Distribution of predicted Tg and bandgap of 12,100 PET-co-(PET-like) copolymers, color-coded by tendency to crystallize. Three known commercial polymers (e.g., PETG, Tritan, Ecozen) reidentified by VFS fall within the desired design region. Structure of labeled polymers are shown in Figs. 3 and 4.

### III. Results and Discussion

Out of designed copolymers, 3,390 cases met the target criteria of Tg higher than PET's predicted Tg (72°C) and bandgap larger than PET's predicted bandgap (3.7 eV), as shown in Fig. 2. Applying a secondary filter based on model confidence, GPR 1-sigma uncertainty of predictions below 30°C for Tg and 0.5 eV for bandgap, narrowed the candidate set to 1,108 polymers. Polymers with a predicted tendency to crystallize equal to or lower than 40% are highlighted in orange-red colormap in Fig. 2. All of 1,108 polymers' tendency to crystalize met <40% criterion. Three commercial copolymers well known as PET replacements, PETG, Tritan, and Ecozen, were among the predicted candidates. These alternatives have been widely adopted due to their improved processability, reduced brittleness, and enhanced clarity compared to PET. Their re-identification through our pipeline demonstrates that such successful PET alternatives can be readily and systematically recovered, providing clear validation of the approach. Figure 3 presents a selection of randomly chosen polymers from the final candidate pool that meet three screening criteria.

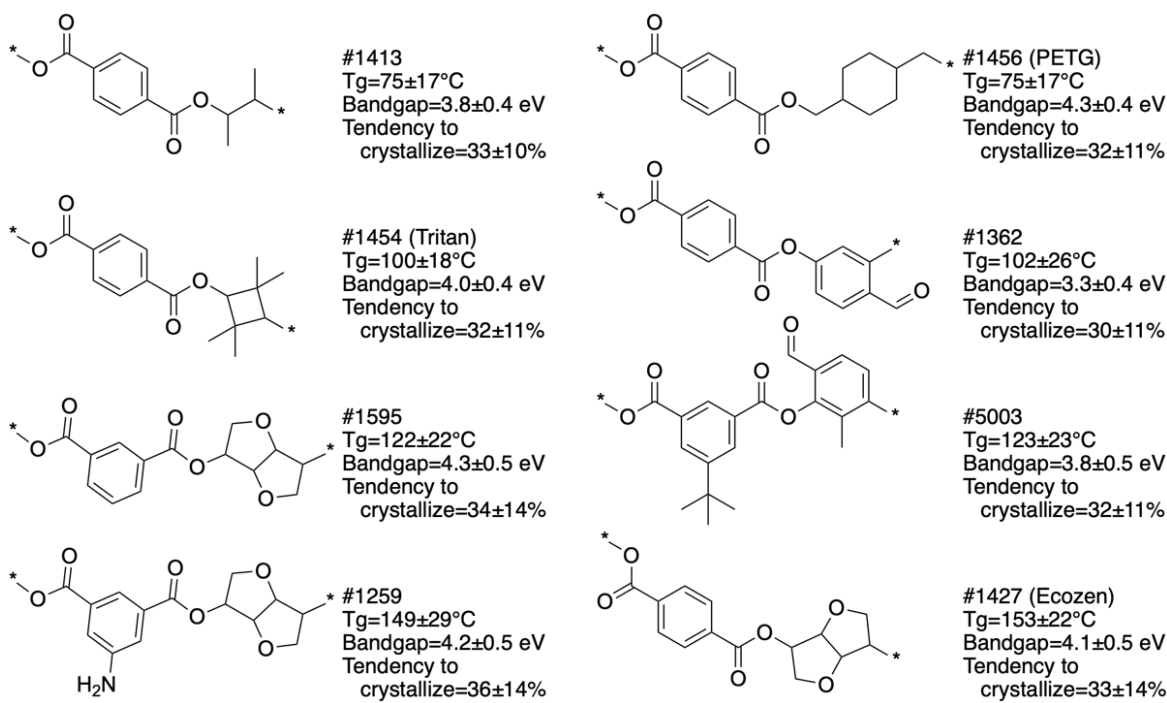

**FIG. 3.** Copolymers potential to replace PET with higher thermal resistance and comparable transparency. Symbols (*) indicate the endpoints of repeating unit. Only PET-like blocks of PET-co-(PET-like) copolymers are visualized.

Taking into account the synthetic accessibility (SA) scores of the PET-like polymers and practical considerations from polymer synthesis experience, two candidates (Fig. 4) were selected for experimental validation. Both polymers were synthesized via solution polycondensation of the diacid chloride (TPA analog) and the diol (EG analog) monomers under basic conditions. The chemical structure and composition of both candidates were confirmed by 1H NMR spectroscopy. (Fig. S1) Sample A exhibited two distinct Tg values, observed at 18.3°C and 101.4°C, respectively. (Fig. S2) The higher Tg notably exceeds that of commercial PET, indicating enhanced thermal rigidity in one phase. The measured Tg of Sample B was 25.1 °C, which is consistent with the predicted Tg value of 33±16°C. The melting temperature (Tm) was observed in the range of 107.1°C - 171.9°C, indicating the presence of semi-crystalline domains. This combination of a low Tg and a broad melting region suggests that Sample B may exhibit rubbery behavior at ambient temperature, making it a promising candidate for use as an elastomeric material. Detailed experimental procedures and additional characterization results are provided in the Supporting Information.

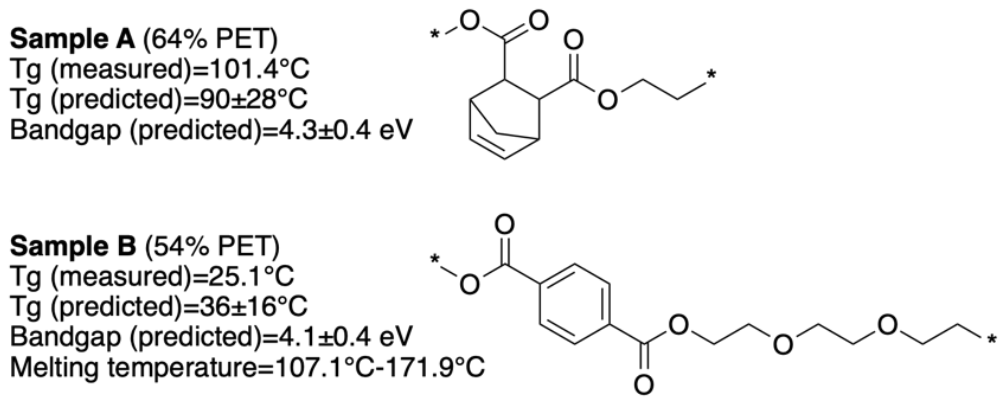

**FIG. 4.** Chemical structure and thermal properties of potential PET replacement copolymers. Symbols (*) indicate the endpoints of repeating unit. Only PET-like blocks of PET-co-(PET-like) copolymers are visualized.

Understanding the relationships between chemical or structural descriptors of designed polymers and their properties provides valuable insights for expanding the design space and further identifying additional novel candidates. Using the Polymer Genome fingerprinting scheme, 283 descriptors, including atomic features, molecular fragments, and extended length-scale characteristics, were generated for 12,100 polymers. SHAP analysis result (Fig. 5) and PCA plots (Fig. S3), highlights the features dominating Tg and bandgap. These include Chi2n (normalized molecular connectivity index for paths of length two), -CH2- (methylene groups not in a ring), and substructures such as C3-C3-O1. Here, C3 and O1 represents a carbon atom bonded to 3 neighboring atoms, and an oxygen atom bonded to 1 neighboring atom.

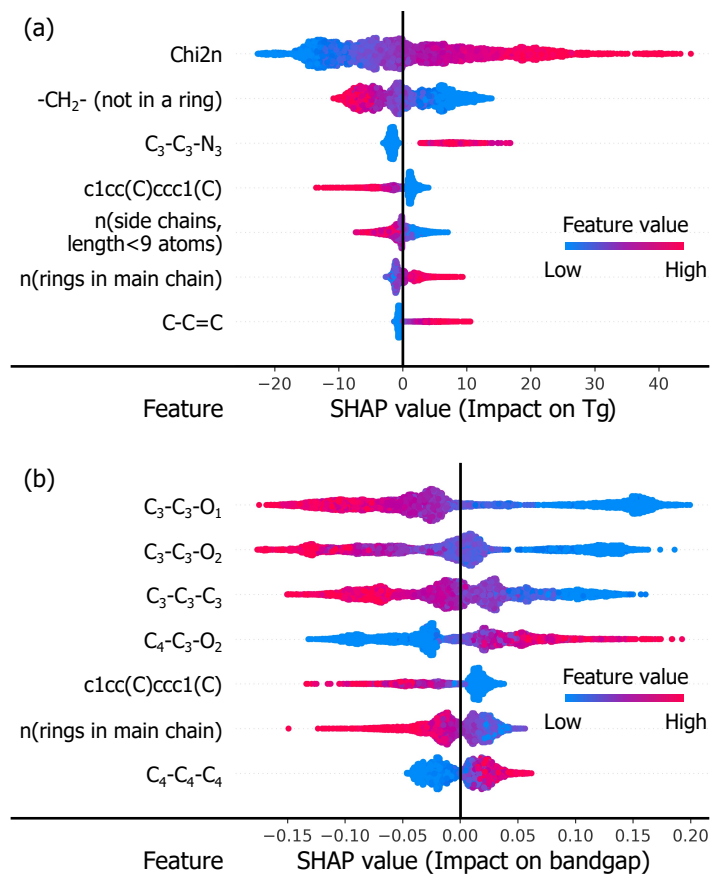

FIG. 5. SHAP analysis showing top 7 most important features influencing (a) Tg and (b) bandgap for 12,100 PET-co-(PET-like) copolymers.

Descriptors such as Chi2n,[31] number of rings in main chain, and C=C double bonds reduce backbone flexibility, contributing to higher Tg, whereas flexible units like -CH2- groups and short side chains lower Tg by enhancing chain mobility. For the bandgap, structural motifs with more $SP^2$-hybridized (three-coordinated) carbons (e.g., C3-C3-O1 and C3-C3-O2) enhances π-electron delocalization, which reduces the bandgap. $SP^3$-rich environments (e.g., C4-C3-O2 and C4-C4-C4) disrupt conjugation and results in a larger bandgap. Fewer aromatic rings composed mainly 3-folded carbons correlate with wider bandgap.

## IV. Conclusions

This proof-of-concept study demonstrates a workflow, generating large number of new polymers inspired by the reaction mechanism of an existing polymer, applying multi-objective screening using ML-predicted properties, and validating synthetic feasibility and performance by lab-test. While commercial PET alternatives with excellent thermal properties already exist, the newly designed copolymers offer promising PET replacements. This study considered only a 1:1 compositional ratio of PET and PET-like units of copolymer. Investigation on broader compositional ranges will help better capturing property trends. The presented framework is generalizable and can be applied to the design of replacements for other polymers. Incorporating practical evaluation factors such as mechanical properties, chemical resistance, toxicity and sustainability, as well as raw material cost and production expenses, would make the study more applicable to real-world scenarios and facilitate the development of scalable, commercially viable designs.

# Supplementary Information

# AI-Driven Design of poly(ethylene terephthalate)-replacement copolymers

Chiho Kim[1,2*], Wei Xiong[1], Akhlak Mahmood[2], Rampi Ramprasad[1,2], Huan Tran[1,2]
1. *School of Materials Science and Engineering, Georgia Institute of Technology, Atlanta, Georgia 30332, USA*
2. *Matmerize, Inc., Atlanta, Georgia 30308, USA*
*E-mail: chiho.kim@gatech.edu

## 1. Experimental validation

### 1.1 Synthesis of Nadic acid chloride

In a typical procedure, an oven-dried 250 mL three-neck round-bottom flask equipped with a 2 cm magnetic stir bar, a 25 mL pressure-equalizing addition funnel, a rubber septum, and an argon inlet was charged with Nadic acid (9.1 g, 50 mmol, 1.0 equiv), anhydrous dichloromethane (DCM) (50 mL), and DMF (0.18 g, 2.5 mmol, 0.05 equiv) as catalyst. The setup was maintained under an inert argon atmosphere throughout the reaction. Oxalyl chloride (13.3 g, 105 mmol, 2.1 equiv) was placed in the addition funnel and added dropwise to the stirred solution over the course of 4 minutes. Vigorous effervescence was observed upon addition. After complete addition, the reaction mixture was stirred at room temperature (23 °C) for an additional 30 minutes. Upon completion, the solvent and excess reagents were removed under reduced pressure using a rotary evaporator, yielding a yellow crude product (acid chloride, ~8.35 g), which was used directly in the subsequent polycondensation step without further purification.

### 1.2 Synthesis of Sample A

All glassware was oven-dried and cooled under a dry nitrogen atmosphere before use. A three-neck round-bottom flask was assembled with a dropping funnel, condenser, and magnetic stir bar, and the system was continuously purged with nitrogen to maintain inert conditions throughout the reaction. Terephthaloyl chloride (TPC) (500 mg, 2.46 mmol, 1.0 equiv) and nadic acid chloride (518 mg, 2.46 mmol, 1.0 equiv) were dissolved together in 10 mL of anhydrous DCM in the dropping funnel. In the reaction flask, ethylene glycol (EG) (306 mg, 4.93 mmol, 2.0 equiv) and triethylamine (TEA) (996 mg, 9.84 mmol, 4.0 equiv) were dissolved in 10 mL of anhydrous DCM and cooled to 0–5 °C using an ice bath. Under vigorous stirring, the mixed acid chloride solution was added dropwise to the cooled EG/TEA solution over 30 minutes, with care taken to maintain the reaction temperature below 10 °C to suppress exothermic effects. After the addition was complete, the reaction mixture was stirred at room temperature (23 °C) for an additional 2 hours. The formation of a white precipitate indicated successful polymer formation. The reaction mixture containing the crude polymer was collected by filtration, washed thoroughly with cold 1 M aqueous HCl to remove residual TEA hydrochloride, and precipitated into excess cold diethyl ether to remove unreacted monomers. Final drying was performed in a vacuum oven at 40 °C for 12 hours, yielding the purified copolyester.

### 1.3 Synthesis of Sample B

All glassware was thoroughly dried in an oven and cooled under a nitrogen atmosphere prior to use. A three-neck round-bottom flask was equipped with a dropping funnel, a condenser, and a magnetic stir bar, and the system was continuously purged with dry nitrogen to maintain an inert atmosphere throughout the reaction. TPC (2.00 g, 9.85 mmol, 2.0 equiv) was dissolved in 10 mL of anhydrous DCM in the dropping funnel. In the reaction flask, triethylene glycol (TEG) (0.74 g, 4.93 mmol, 1.0 equiv), EG (0.31 g, 4.93 mmol, 1.0 equiv), and TEA (1.99 g, 19.7 mmol, 4.0 equiv) were dissolved in 10 mL of anhydrous DCM. The resulting glycol/TEA solution was cooled to 0–5 °C using an ice-water bath. With vigorous stirring, the TPC solution was added dropwise to the cooled glycol/TEA solution over a period of 30 minutes, ensuring that the reaction temperature was maintained below 10 °C to minimize exothermic effects. Upon completion of the addition, the reaction mixture was allowed to warm to room temperature (~23 °C) and stirred for an additional 2 hours. The formation of a white precipitate was observed during the reaction, indicating successful polymer formation. The reaction mixture containing the crude polymer was collected by filtration, washed thoroughly with cold 1 M aqueous HCl to remove residual TEA hydrochloride, and precipitated into excess cold diethyl ether to remove unreacted monomers. The solid polymer was dried under reduced pressure using a rotary evaporator and further dried in a vacuum oven at 40 °C for 12 hours to obtain the final purified product.

**1.4 Characterization**

The chemical structure and composition of the synthesized copolymers were confirmed by **¹H nuclear magnetic resonance (¹H NMR) spectroscopy** in **CDCl₃** at room temperature. **Characteristic peaks corresponding to both monomer units were clearly observed in the ¹H NMR spectra (Figure S1). Quantitative integration** of representative signals revealed that the molar ratio of the two ester units was approximately **1.79:1** in **Sample A** and **1.16:1** in **Sample B**, respectively. The glass transition temperature (Tg) was measured using differential scanning calorimetry (DSC) under a nitrogen atmosphere with a heating rate of 10 °C/min. (Figure S2)

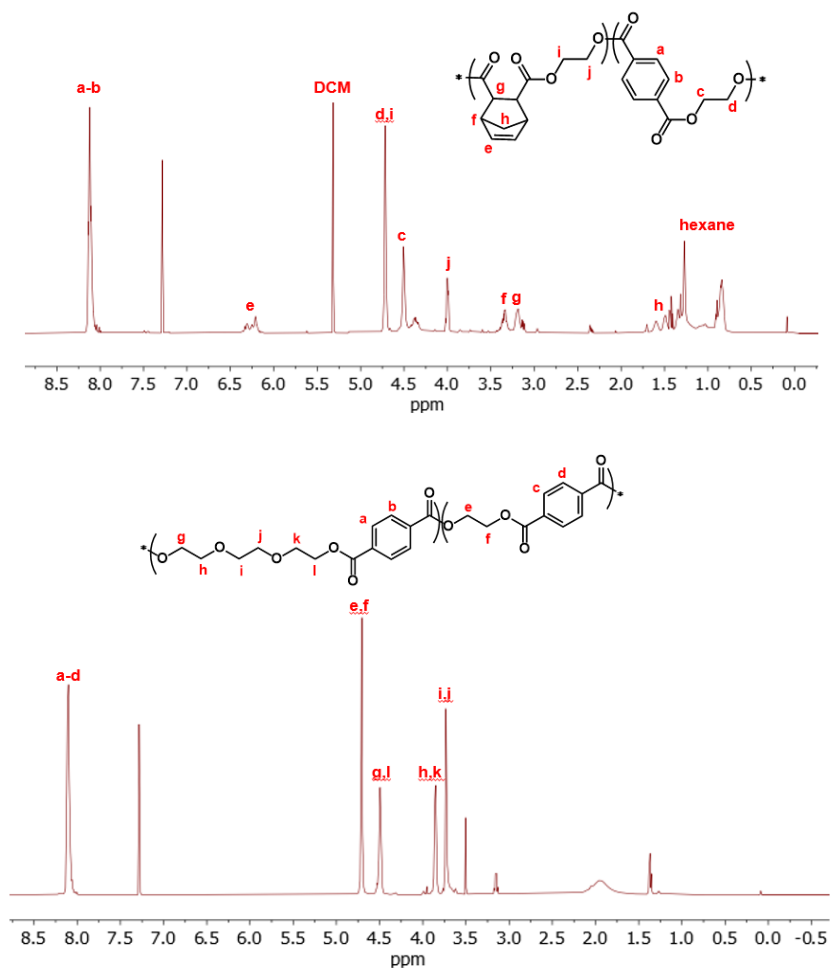

**FIG. S1.** ¹H NMR spectra of Sample A (top) and Sample B (bottom).

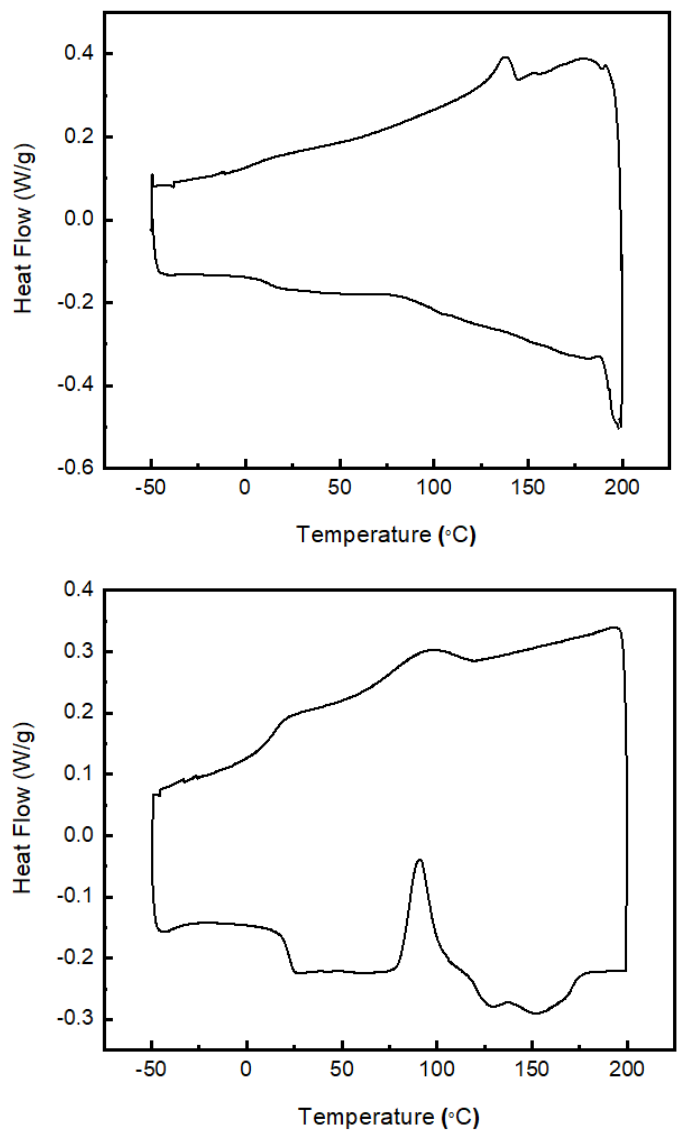

**FIG. S2.** DSC curves of Sample A (top) and Sample B (bottom).

## 2. Pattern analysis

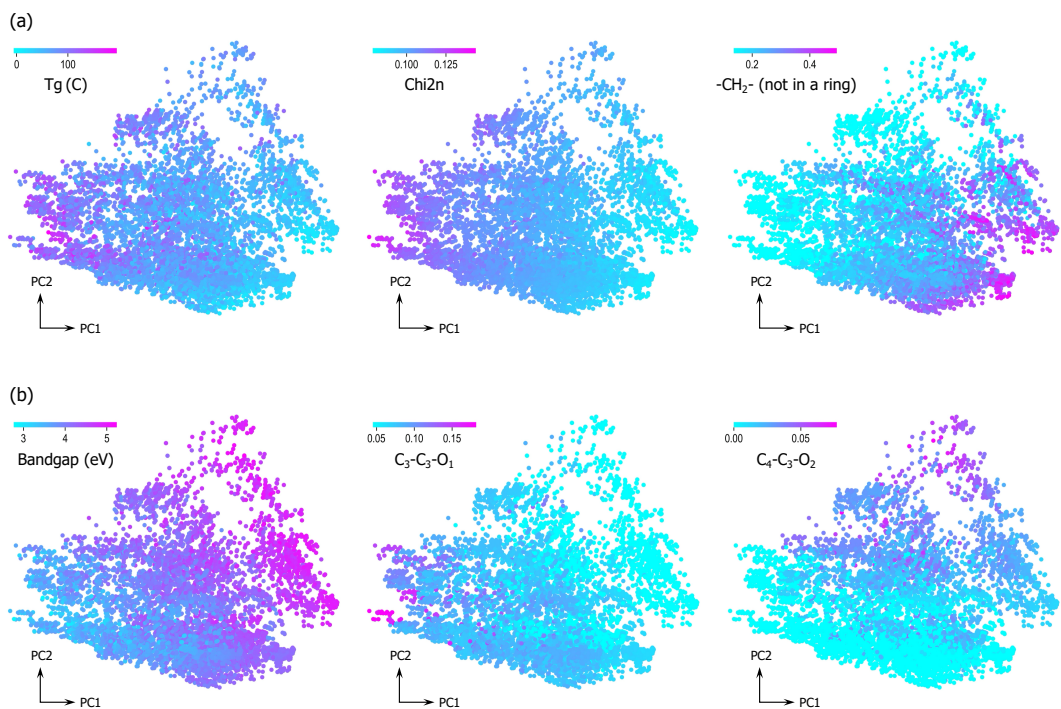

**FIG. S3**. Visualizations of the 12,100 PET-co-(PET-like) copolymers projected onto a 2D space spanned by PC1 and PC2, two first principal axes obtained by a PCA. Color bars are used for encoding 1) the property values, Tg in (a) and bandgap in (b), and 2) two descriptors with largest impact on each property in positive and negative direction.

19